\documentclass[aps,prd,preprint,tightenlines,superscriptaddress]{revtex4-1}
\usepackage{epsfig}
\usepackage{graphicx}
\usepackage{psfrag}
\usepackage{amsmath,amssymb}
\usepackage{colordvi}
\usepackage{color}
\usepackage{amsfonts}
\usepackage{enumerate}
\usepackage{slashed,bm}

\def \non {\nonumber}
\def \beq  {\begin{equation}}
\def \eeq  {\end{equation}}

\begin{document}

\title{Gaussian-weighted Parton Quasi-distribution}

\collaboration{\bf{Lattice Parton Physics Project ($\rm {\bf LP^3}$)}}


\author{Tomomi Ishikawa}
\affiliation{T.~D.~Lee Institute, Shanghai Jiao Tong University, Shanghai, 200240, P. R. China}

\author{Luchang Jin}
\affiliation{Physics Department, Brookhaven National Laboratory, Upton, New York 11973, USA}

\author{Huey-Wen Lin}
\affiliation{Department of Physics and Astronomy, Michigan State University, East Lansing, MI 48824}
\affiliation{Department of Computational Mathematics, Michigan State University, East Lansing, MI 48824}

\author{Andreas Sch\"afer}
\affiliation{Institut f\"ur Theoretische Physik, Universit\"at Regensburg, D-93040 Regensburg, Germany}

\author{Yi-Bo Yang}
\affiliation{Department of Physics and Astronomy, Michigan State University, East Lansing, MI 48824}

\author{Jian-Hui Zhang}
\affiliation{Institut f\"ur Theoretische Physik, Universit\"at Regensburg, D-93040 Regensburg, Germany}

\author{Yong Zhao}
\affiliation{Center for Theoretical Physics, Massachusetts Institute of Technology, Cambridge, MA 02139, USA
}
\vspace{2in}

\preprint{MSUHEP-17-021}

\begin{abstract}
We propose a revised definition of quasi-distributions within the framework of large-momentum effective theory (LaMET) that improves convergence towards the large-momentum limit. Since the definition of quasi-distributions is not unique, each choice goes along with a specific matching function, we can use this freedom to optimize convergence towards the large-momentum limit. As an illustration, we study quasi-distributions with a Gaussian weighting factor that naturally suppresses long-range correlations, which are plagued by artifacts. This choice has the advantage that the matching functions can be trivially obtained from the known ones. We apply the Gaussian weighting to the previously published results for the nonperturbatively renormalized unpolarized quark distribution, and find that the unphysical oscillatory behavior is significantly reduced. 
\end{abstract}

\maketitle


Recent years have witnessed rapid progress in calculating parton distribution functions (PDFs), rather than just their moments, from lattice QCD~\cite{Ji:2013dva,Xiong:2013bka,Lin:2014zya,Ma:2017pxb,Alexandrou:2015rja,Chen:2016utp,Alexandrou:2016jqi,Chen:2017mzz,
Alexandrou:2017dzj}. The most remarkable aspect of such calculations is that one can connect appropriately chosen Euclidean correlation functions calculable on the lattice to PDFs through a perturbative matching relation. This can be understood in the framework of the large-momentum effective theory (LaMET)~\cite{Ji:2014gla,Ji:2017rah}. 
According to LaMET, one can construct appropriate quasi-PDFs which are time-independent matrix elements defined at finite hadron momentum, and approach the PDFs in the infinite momentum limit. The PDFs can be extracted from lattice calculations of quasi-PDFs through a perturbative matching, up to power corrections suppressed by the hadron momentum. It is worthwhile to stress that the choice of quasi-PDFs is not unique; any construction is allowed as long as it approaches the PDFs in the infinite momentum limit and is practically calculable. This provides a flexibility to optimize the PDF results by choosing a suitable quasi-PDF that is best suited for lattice simulations.

The LaMET approach has been applied to compute the nucleon unpolarized, helicity and transversity PDFs~\cite{Lin:2014zya,Alexandrou:2015rja,Alexandrou:2016jqi,Chen:2016utp,Chen:2017mzz}, as well as the pion distribution amplitude (DA)~\cite{Zhang:2017bzy}. 
A first lattice PDF calculation at physical pion mass has also recently become available~\cite{Lin:2017ani} (see also~\cite{Alexandrou:2017dzj}). 
In addition to LaMET, there also exist other proposals to use similar matrix elements or current-current correlators at spacelike separation to compute PDFs, meson DAs, etc.~\cite{Davoudi:2012ya,Detmold:2005gg,Liu:1993cv,Liang:2017mye,Braun:2007wv,Radyushkin:2017cyf,Bali:2017gfr,Ma:2017pxb}.

The definition of quasi-PDFs in LaMET involves a spacelike Wilson line, which gives rise to linear divergences that have to be nonperturbatively renormalized. As shown in Refs.~\cite{Dorn:1986dt,Ishikawa:2016znu,Chen:2016fxx,Ji:2017oey,Ishikawa:2017faj,Green:2017xeu}, the renormalization of  such linear divergences takes an exponential form that enhances the long-range correlations, and thereby significantly increases the uncertainties of the result~\cite{Chen:2017mzz}. 
A large value of the product of the hadron momentum and the length of the Wilson line, or the Ioffe-time, $z p_z$, is responsible for the small-$x$ behavior of the PDF when Fourier transformed to momentum space.
Given the small lattice hadron momentum used in current lattice calculations, the extracted PDF can not be trusted in the $x \lesssim 0.3$ region. In addition, since the matching is done perturbatively, using long distance input to reach large $z p_z$ is also not justified, i.e. it introduces large systematic uncertainties. In, e.g.~\cite{Radyushkin:2017cyf} such contributions are suppressed by construction, forcing PDFs to vanish for Bjorken $x>1$. In the LaMET approach we chose not to do so to be able to use the obtained large Bjorken $x$ behavior as handle to estimate systematic uncertainties. The latter is needed if such results are used in global PDF analysis as in~\cite{Nocera:2017war}. Unless $p_z$ can be made large enough to make the systematic errors small, lattice-PDFs will carry only very little weight in such fits. In~\cite{Lin:2017ani} two filtering techniques were thus used to estimate how large a $p_z$ has to be reached to obtain reliable PDFs in the LaMET approach, where they also account for the truncation of the long-range correlations. To estimate the systematics introduced by the filter function, the authors of Ref.~\cite{Lin:2017ani} proposed to study the parameter setup done in actual lattice calculations using a known PDF input transformed to coordinate space and then back to $x$-space. This allows a very rough estimation of
the systematics due to the Fourier transformation, the smallness of the
hadron momentum and the lattice finite volume. 

Here we propose a more systematic way to achieve the same effect.
Just as there are many lattice fermion actions which lead to the same continuum Dirac fermion action, there are also a vast number of quasi-PDF constructions on the lattice which can reach the true lightcone PDF after the proper matching. 
We choose to revise Ji's quasi-PDFs definition by including a Gaussian weighting factor that suppresses long-range correlations and removes the unphysical oscillatory behavior. This construction only introduces a single parameter that we can control systematically with a known asymptotic large-momentum limit.  Furthermore, all the perturbative matching results already done for the original quasi-PDF can be straightforwardly converted to those with the revised quasi-PDFs. 
As a demonstration, we show the impact of the Gaussian weighting on the isovector unpolarized quark PDF result presented in Ref.~\cite{Chen:2017mzz}.\vspace*{1em}

Let us consider the definition of the quasi-PDF, taking the unpolarized quark quasi-PDF as an example. The discussion below applies equally well to other quark or gluon quasi-PDFs, meson quasi-DAs, etc. The unpolarized quark quasi-PDF is defined as
\beq\label{quark-PDF}
   \tilde q(x, \Lambda, p_z) = \int \frac{dz}{4\pi} e^{ix p_z z}  \langle p|\overline{\psi}(0, 0_\perp, z)
   \gamma^z L(z,0)\psi(0) |p\rangle \ ,
\eeq
where the quark fields are separated along the spatial $z$-direction, $L(z,0)$ is the Wilson-line gauge link inserted to ensure gauge invariance, $p_\mu=(p_0,0,0,p_z)$ is the nucleon momentum, and $\Lambda$ denotes the UV cutoff or renormalization scale in an appropriate scheme such as the RI/MOM scheme used in Refs.~\cite{Chen:2017mzz,Stewart:2017tvs}.

Ref.~\cite{Ji:2013dva} showed that the above quark quasi-PDF approaches the normal quark PDF in the infinite-momentum limit $p_z\to\infty$. It is worthwhile to point out that, although the $z$-integration in Eq.~(\ref{quark-PDF}) goes over non-perturbative large distances, for a large $p_z$ the effective integration range shrinks to small distances $z\sim 1/p_z$. In other words, the large $p_z$ limit ensures the smallness of higher-twist corrections as well as perturbative corrections. In contrast, in the pseudo-distribution this is achieved by explicitly keeping $z^2$ small. Nevertheless, a large momentum is still required to probe information on higher moments. Therefore, the quasi- and pseudo-distribution approaches are actually equivalent. 

In the following we will focus on the quasi-PDF and consider our revised definition for it. As mentioned earlier, any quark quasi-PDF that becomes the normal PDF in the limit $p_z\to\infty$ and is practically calculable offers an equally good candidate for lattice simulations. We consider the following definition with a Gaussian weighting factor
\beq\label{gaussian-PDF}
   \tilde q_\text{GW}(x, \Lambda, p_z) = \int \frac{dz}{4\pi} e^{ix p_z z- z^2/l^2}  \langle p|\overline{\psi}(0, 0_\perp, z)
   \gamma^z L(z,0)\psi(0) |p\rangle \ ,
\eeq
where $l$ denotes a length scale. The above definition is related to the original one in Eq.~(\ref{quark-PDF}) by
\begin{align}\label{qGandqrelation}
\tilde q_\text{GW}(x,\Lambda, p_z)&=\int\frac{dz}{4\pi}e^{i x p_z z}\Big[2p_z\int dy e^{-i y p_z z}\tilde q(y, \Lambda, p_z)\Big]\int dt \frac{l p_z}{2\sqrt{\pi}}e^{-\frac{t^2 l^2 p_z^2}{4}}e^{-i t p_z z}\non\\
&=\int dy\, \tilde q(y, \Lambda, p_z) \Big[\frac{l p_z}{2\sqrt{\pi}}e^{-\frac{l^2(x-y)^2 p_z^2}{4}}\Big].
\end{align}
In the limit $l p_z\gg 1$, the term in the square bracket yields a $\delta$-function $\delta(x-y)$ and $\tilde q_\text{GW}$ becomes $\tilde q$. For a relatively large $l p_z$, $\tilde q_\text{GW}$ differs from $\tilde q$ only by power corrections suppressed by $1/(l p_z)^2$. Therefore, $\tilde q_\text{GW}(x)$ is an equally good candidate for the quasi-PDF. In practice, the weighting parameter $l$ can be tuned to optimize the PDF result extracted from lattice simulations, provided that the additional errors introduced by the weighting are under control.

In realistic lattice simulations, one can only have a finite lattice size and nucleon momentum. As a consequence, the $z$-integration in Eq.~(\ref{quark-PDF}) is always truncated, which then leads to an unphysical oscillation in $\tilde q$ obscuring the physical information embedded in it. This has been seen in the result of Ref.~\cite{Chen:2017mzz}. As we will see below, Eq.~(\ref{gaussian-PDF}) has the advantage that, with an appropriate choice of the weighting parameter, the unphysical oscillation can be removed while the physical result remains unchanged in the moderate-to-large Bjorken $x$ region. The price one has to pay is that the result in the small Bjorken $x$ region is changed by the weighting. This is the most difficult region for lattice simulations and can not be trusted anyway at the current stage.

\begin{figure}[tbp]
\centering
\includegraphics[width=0.49\textwidth]{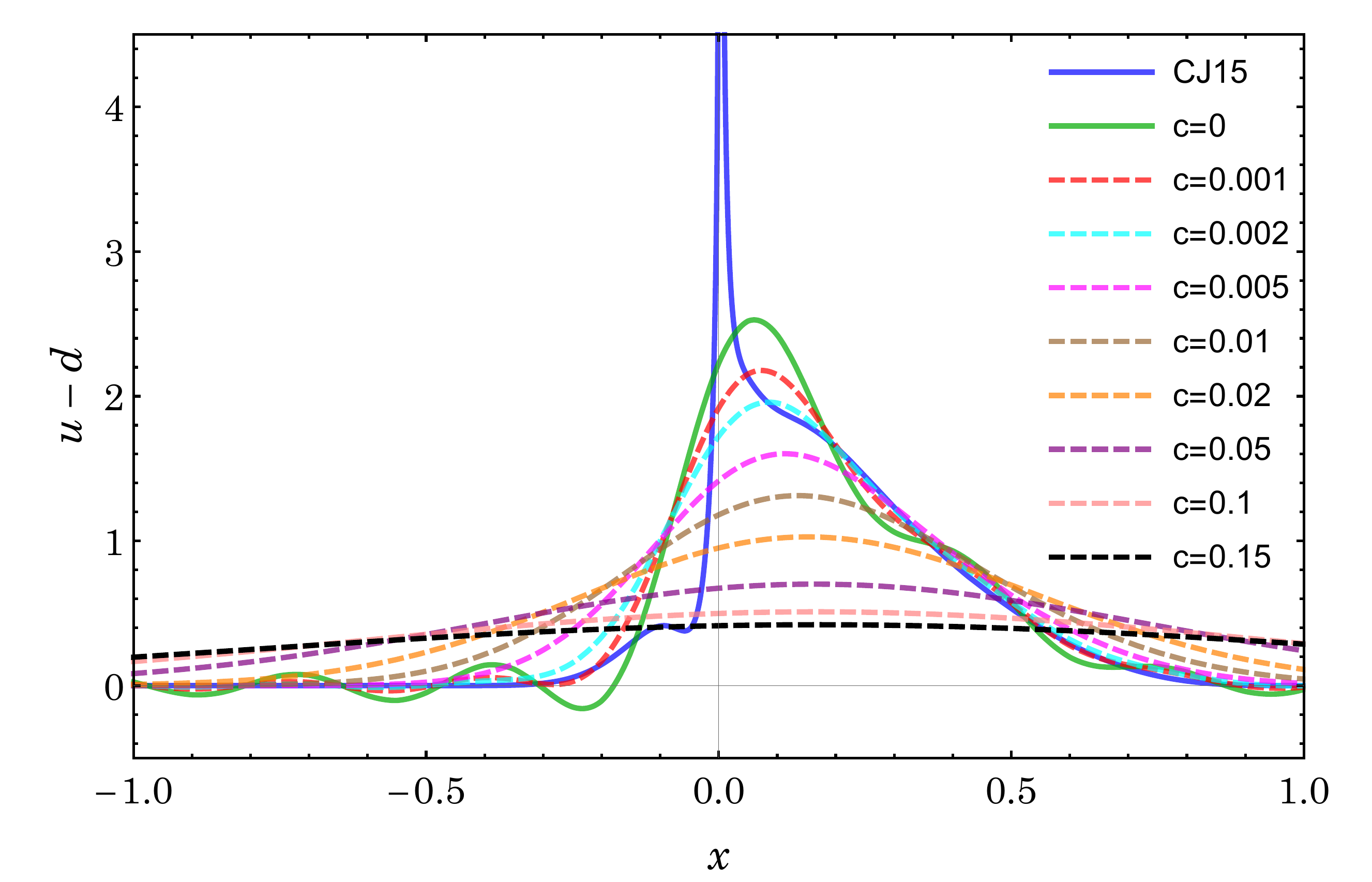}
\includegraphics[width=0.49\textwidth]{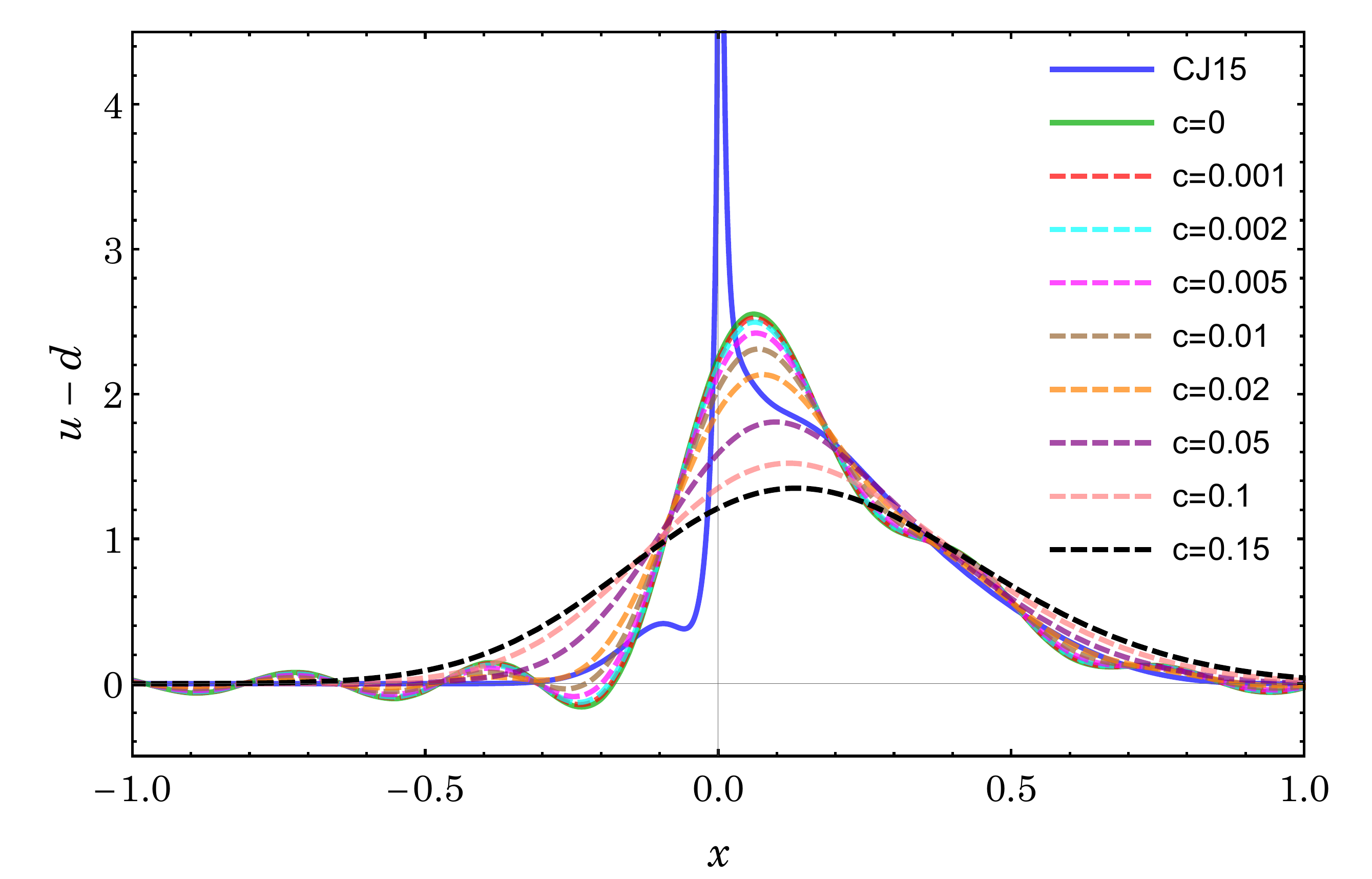}
\includegraphics[width=0.49\textwidth]{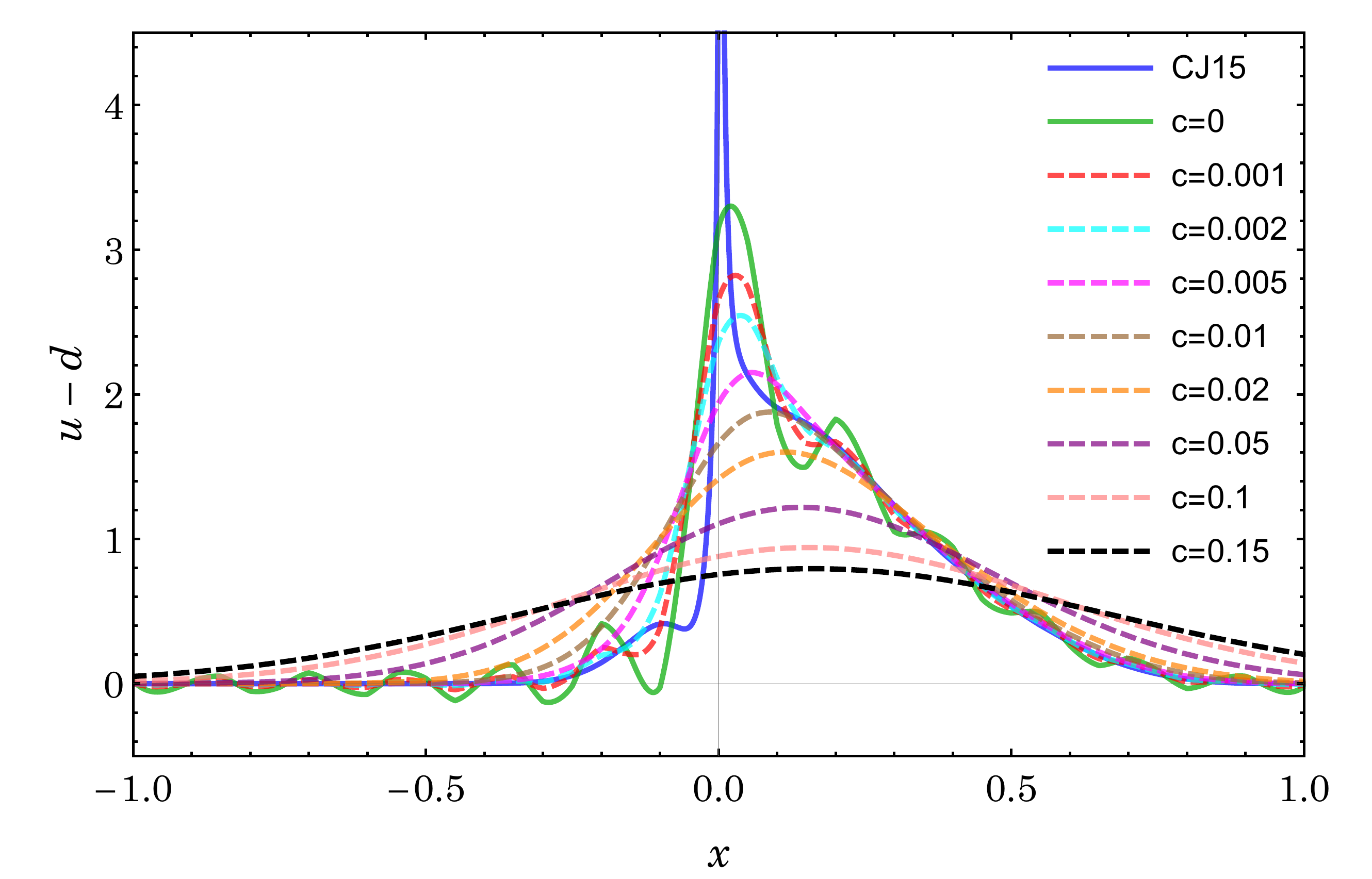}
\includegraphics[width=0.49\textwidth]{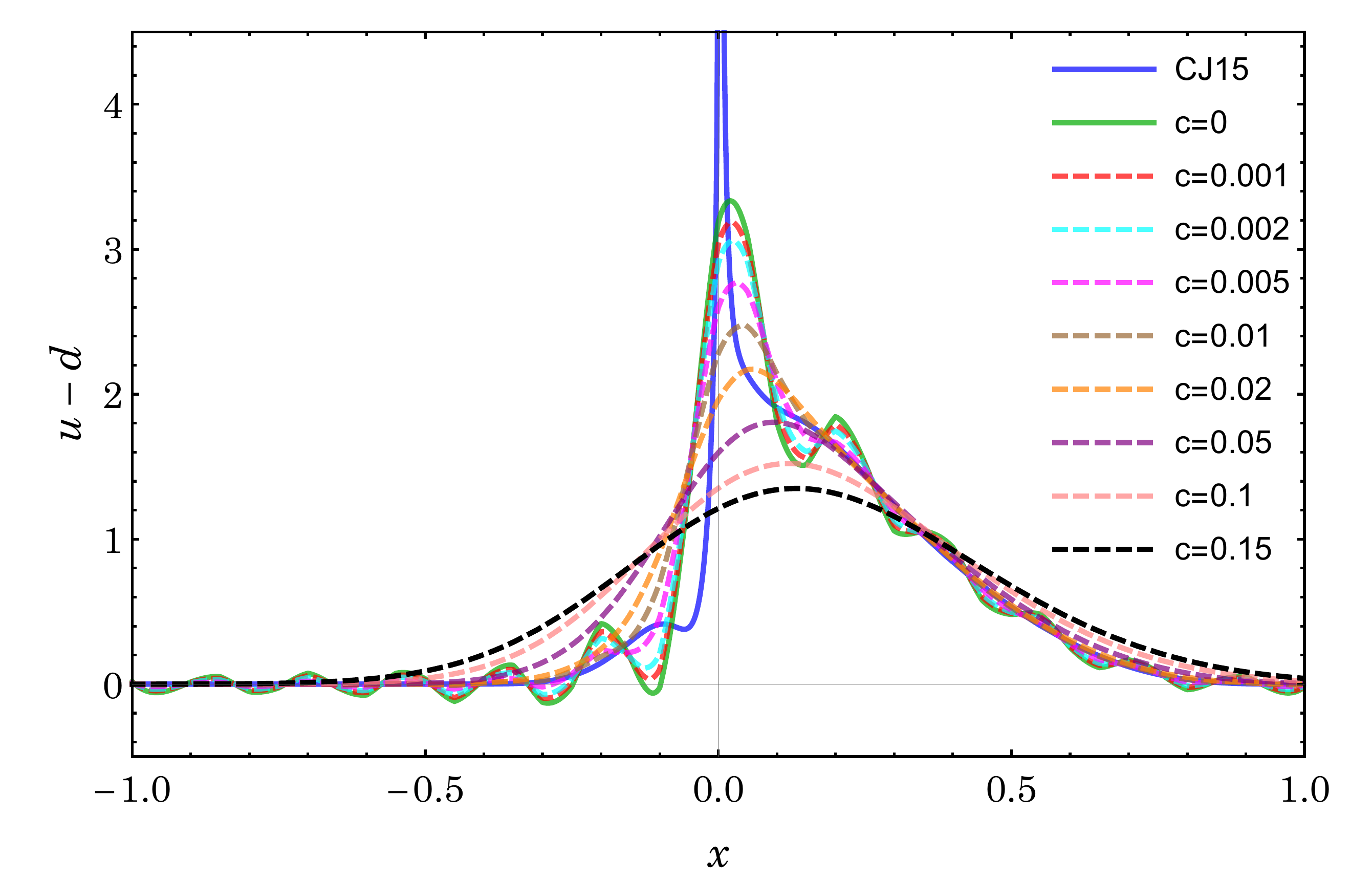}
\includegraphics[width=0.49\textwidth]{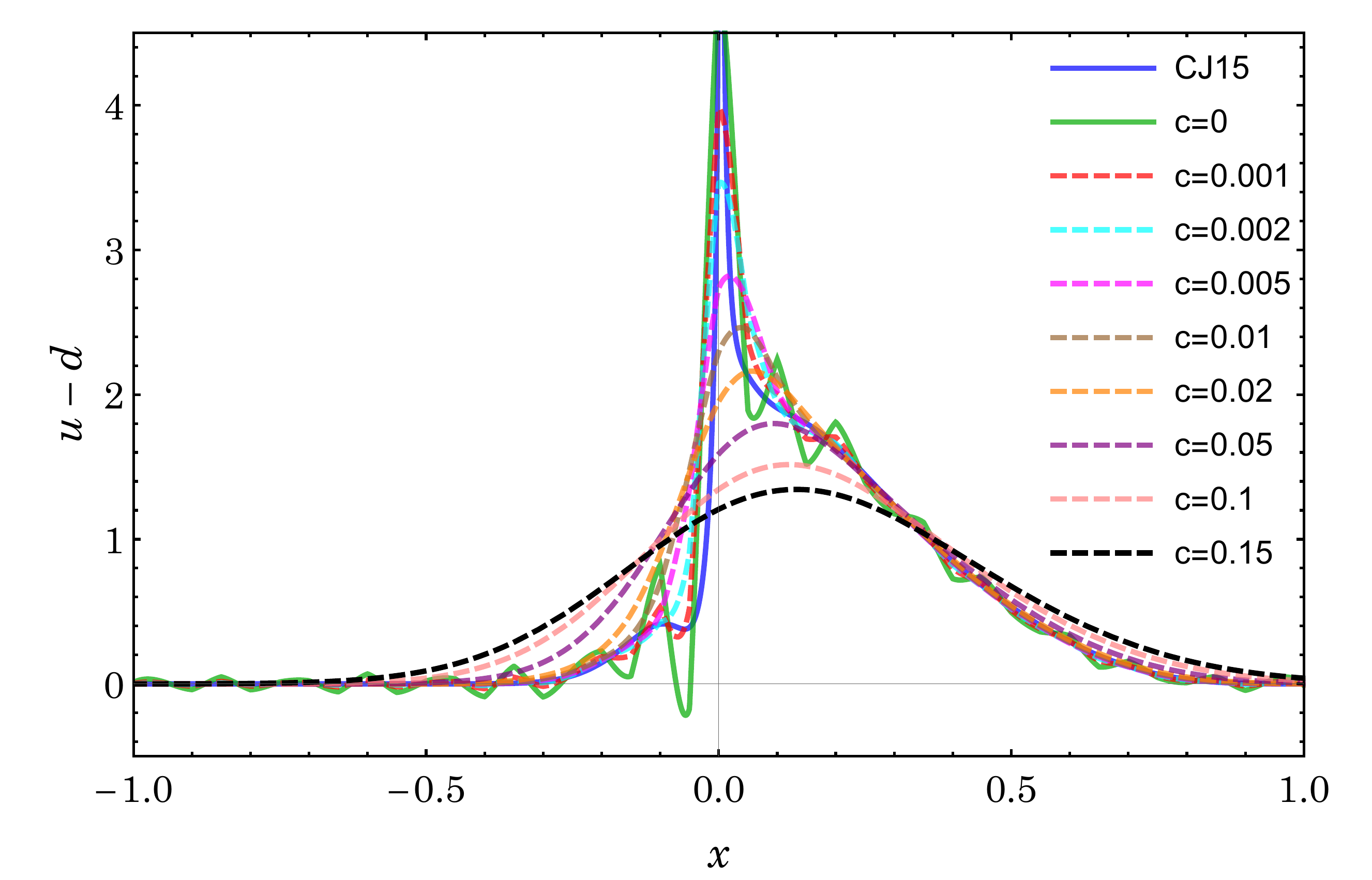}
\includegraphics[width=0.49\textwidth]{pdf_gaussian_weighting_pz24z32.pdf}

\caption{CJ15 PDF with different Gaussian weighting. The blue curve is the original PDF, ``c=0" denotes the Fourier transform result of a truncated coordinate space matrix element generated from CJ15 data without weighting. The dashed curves correspond to different Gaussian weighting factors. The plots correspond to $a p^z=\frac{2\pi}{64}\{6, 12, 24\}$ with $|z_{\rm{max}}|/a=32$ (left column from top to bottom) and $a p^z=\frac{2\pi}{64}\times 24$ with $|z_{\rm{max}}|/a=\{8, 16, 32\}$ (right column from top to bottom), respectively.}
\label{cj15comparison}
\end{figure}

Some general remarks on the choice of weighting parameter follow. On one hand, in order that power corrections of $\mathcal O(1/(l p_z)^2)$ can be safely neglected, one needs $l p_z\gg 1$. On the other hand, for a given $p_z$, $z^2/l^2$ shall not be too small for large $z$ in order to have a suppression of the long-range correlations. To find an optimal choice of $l$, let us take the CJ15 PDF~\cite{Accardi:2016qay} as an example. For simplicity, we use the central value only. We first Fourier transform the CJ15 PDF to coordinate space, then implement the Gaussian weighting in a truncated region, and transform back to momentum space. The Gaussian weighting is chosen to be in the form of $e^{-c z^2/a^2}$ with $a$ the lattice spacing and $c=a^2/l^2$. The results are shown in Fig.~\ref{cj15comparison}. The green curves correspond to the Fourier transform of a truncated coordinate-space matrix element (with $a p_z=2\pi/64\{6, 12, 24\}$ and $|z|/a\le32$ for the left column, and $a p_z=2\pi/64\times 24$ with $|z|/a\le\{8, 16, 32\}$ for the right column) generated from the CJ15 PDF data. A truncation like this is unavoidable in a realistic lattice simulation, and leads to an oscillation around the original CJ15 curve. As can be seen from the figure, adding a weighting factor indeed improves the oscillatory behavior. For a relatively large $p_z$ in the right column, there exists a wide range of choices for $c$ that removes the unphysical oscillation while retaining the original physical distribution at moderate to large $x$. This requires $|z|p_z$ to reach at least $\sim20$ (estimated from the top right panel of Fig.~\ref{cj15comparison} with $z_{\rm max}\,p_z= 32a\times 12\pi/(64a)\sim 20$). Moreover, the weighted distribution in the moderate-to-large $x$ region remains essentially unchanged from top to bottom in the right column, indicating that information from large $z$ is not important for constructing the PDF in that region. Of course, the distribution in small-$x$ region changes significantly since it is determined by long-range correlations. The left column shows that the larger momentum one has, the larger regions of the original curve one can recover from the weighted distribution and the less sensitivity to $c$ one has at moderate to large $x$. In general, it is optimal to choose $l$ in the perturbative range, in order to keep control of non-perturbative contributions. If we have $a=0.05\,\mbox{fm}$, then for $c=0.05$ we have $l\sim 1\, \mbox{GeV}^{-1}$, the weighted curve still deviates from the original curve at moderate to large $x$ in the second plot of the left column, while it essentially coincides with the original curve in the same region in the last plot of the left column, which has twice the hadron momentum. This indicates that a larger momentum is important to achieve a suppression of unphysical oscillation without changing physics at moderate to large $x$.

\begin{figure}[tbp]
\centering
\includegraphics[width=0.49\textwidth]{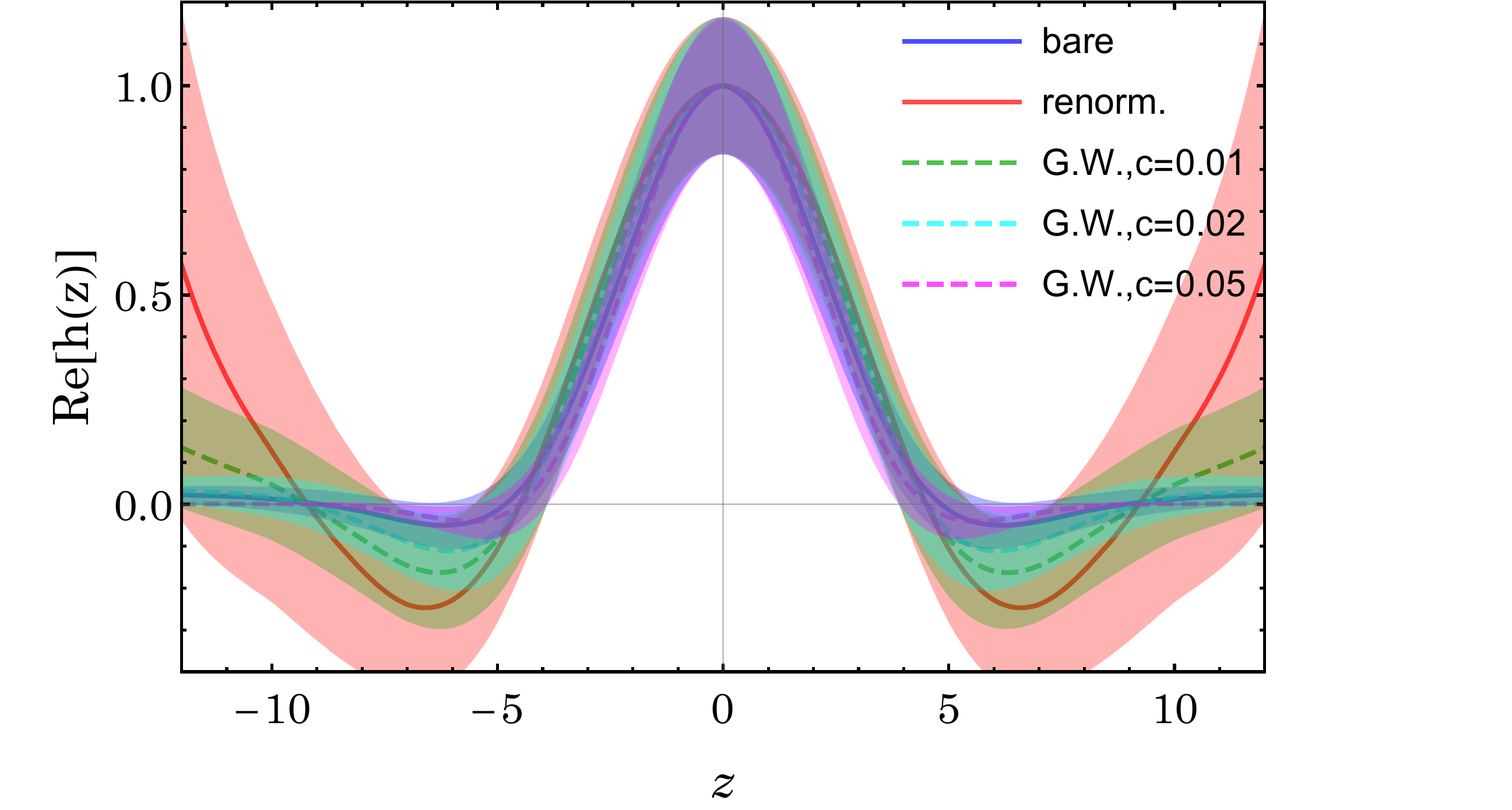}
\includegraphics[width=0.49\textwidth]{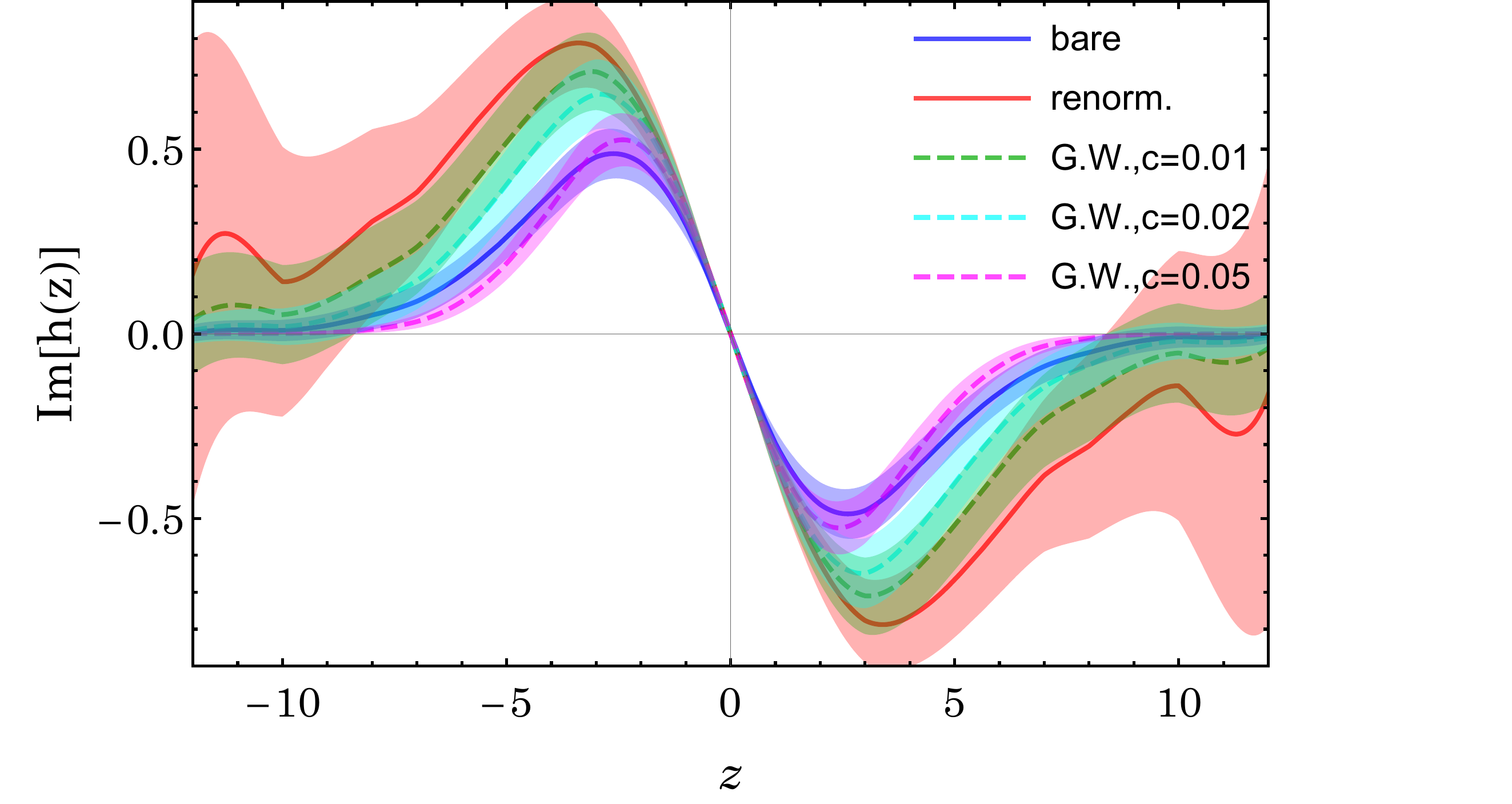}
\caption{Impact of the Gaussian weighting factor on lattice matrix elements. The left plot shows the real part, and the right the imaginary part of the matrix elements.}
\label{gaussianME}
\end{figure}

The above example indeed gives some hint on the appropriate choice of $c$. In the following, we apply the Gaussian weighting to the lattice data presented in Ref.~\cite{Chen:2017mzz}, based on the above observation. Fig.~\ref{gaussianME} shows the bare and renormalized matrix elements for nucleon momentum $p^z=6\pi/L$ with $L\approx 3\, fm$, together with the Gaussian weighted renormalized matrix elements. As an illustration, we show in Fig.~\ref{gaussianME} the Gaussian weighted results for $c=0.01, 0.02, 0.05$, respectively. One can clearly see a suppression of long-range correlations after including the Gaussian weighting factor.

From Eq.~(\ref{qGandqrelation}), it is straightforward to write down the matching for $\tilde q_{GW}$, which is related to the matching for $\tilde q$ as following
\begin{align}\label{mommatching}
&\tilde q_{GW}(x, \Lambda, p^z)\non\\
&=\int_{-1}^1\frac{du}{|u|}\int dy\, Z\Big(\frac{y}{u},\frac{\Lambda}{u p^z},\frac{\mu}{u p^z}\Big)q(u, \mu)\Big[\frac{l p^z}{2\sqrt{\pi}}e^{-\frac{(x-y)^2l^2 p_z^2}{4}}\Big]+\mathcal O(\Lambda_{\rm{QCD}}^2/p_z^2,M^2/p_z^2,1/(l p^z)^2),\non\\
&=\int_{-1}^1\frac{du}{|u|}Z\Big(\frac{x}{u},\frac{\Lambda}{u p^z},\frac{\mu}{u p^z}\Big)q(u, \mu)+\mathcal O(\Lambda_{\rm{QCD}}^2/p_z^2,M^2/p_z^2,1/(l p^z)^2),
\end{align}
where the power corrections $\mathcal O(\Lambda_{\rm{QCD}}^2/p_z^2,M^2/p_z^2, 1/(l p^z)^2)$ in the first line are related to those of the original quasi-distribution by Eq.~(\ref{qGandqrelation}), while in the second line we move the additional $\mathcal O(1/(l p^z)^2)$ power correction from introducing the weighting to the second term. The extra $\mathcal O(1/(l p^z)^2)$ power correction introduced by Gaussian weighting is similar to that introduced by the gradient flow in Ref.~\cite{Monahan:2016bvm}, where one needs a non-zero flow time $\tau$ to ensure a finite continuum limit of the quasi-PDF, but also needs to keep $\tau$ small to suppress the extra $\mathcal O(\sqrt{\tau}\Lambda_{\rm{QCD}})$ corrections. It is therefore important to find a window for $\tau$ so that both can be achieved. In our case, we need to keep a non-zero $1/l$ to suppress long-range correlations, on the other hand, $1/l$ shall not be too large so that the extra $\mathcal O(1/(l p^z)^2)$ corrections are still under control.

\begin{figure}[tbp]
\centering
\includegraphics[width=0.7\textwidth]{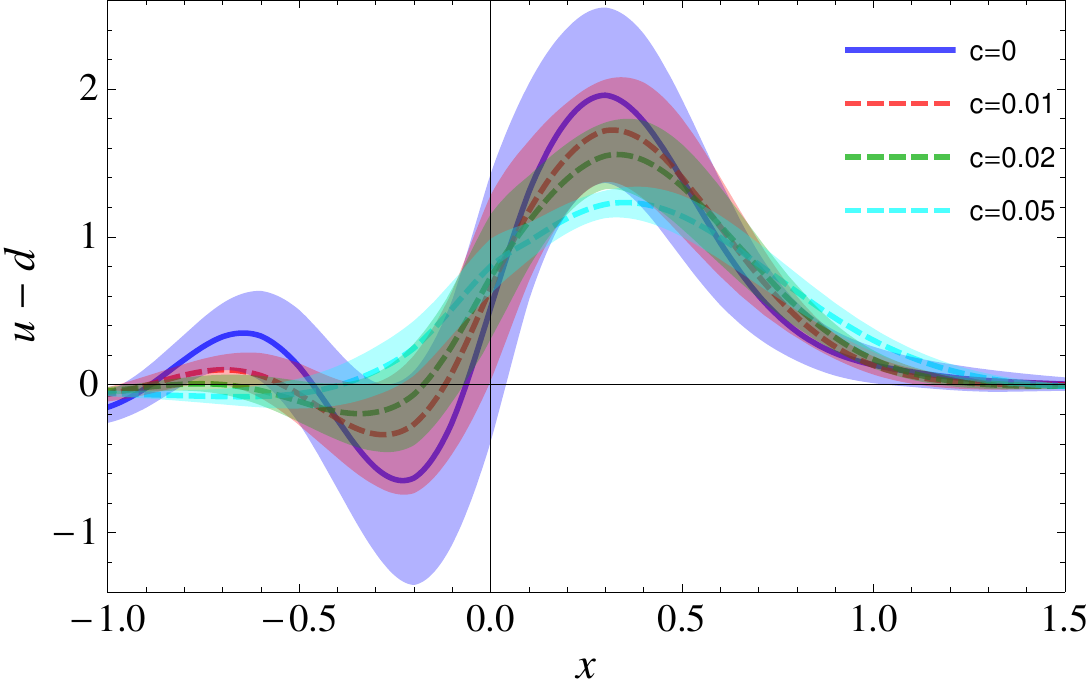}
\caption{Impact of the Gaussian weighting factor on momentum space distribution.}
\label{gaussianfig}
\end{figure}

In Fig.~\ref{gaussianfig}, we show the impact of the Gaussian weighting on momentum-space distribution obtained in Ref.~\cite{Chen:2017mzz}. To estimate the extra $\mathcal O(1/(l p_z)^2)$ errors in Eq.~(\ref{mommatching}), we did a simple extrapolation to infinite momentum as in Ref.~\cite{Chen:2017mzz}, {\it i.e.} we have assumed a simple functional form of the power corrections like $a(x)+b(x)/p_z^2$ and used the results at different $p_z$ to determine the leading term $a(x)$. As can be seen from the plot, including the Gaussian weighting reduces the oscillation in the negative $x$ region observed in Ref.~\cite{Chen:2017mzz}, which was due to a cutoff in long-range correlations. Moreover, the Gaussian weighting introduces a significant change for the distribution in the small-$x$ region, but only a slight change at moderate to large $x$, reflecting the large systematic error in the small-$x$ region for current lattice setups. The maximum $zp_z$ in this study is obviously not large enough, but the results provide a quantitative estimate on which $x$ region the PDF extracted from current lattice data can be trusted.

\vspace*{1em}

In this paper, we have considered a revised quasi-distribution by including a Gaussian weighting factor, which naturally suppresses the long-range correlation, even after renormalization of the Wilson line. It improves the oscillatory behavior of the isovector unpolarized quark PDF observed in Ref.~\cite{Chen:2017mzz}. The advantage of the revised quasi-PDF considered in this paper is that the matching function and power corrections can be straightforwardly obtained from the known ones for the standard quasi-PDF. A straightforward generalization of the Gaussian weighting presented in this paper is to use a superposition of different Gaussian factors or a $p_z$-dependent weighting parameter $c$, which can also be easily implemented. It is highly probable that more effective modifications of quasi-PDFs can be found, which will, however, usually require a recalculation of the matching function and power corrections. \vspace*{1em}


{\it We thank Jiunn-Wei Chen for useful discussions. TI is supported by Science and Technology Commission of Shanghai Municipality (Grants No. 16DZ2260200). TI and LCJ are supported by the Department of Energy, Laboratory Directed Research and Development (LDRD) funding of BNL, under contract DE-EC0012704. The work of HL and YY is supposed by US National Science Foundation under grant PHY 1653405. AS and JHZ are supported by the SFB/TRR-55 grant ``Hadron Physics from Lattice QCD'', JHZ is also supported by a grant from National Science Foundation of China (No.~11405104). YZ is supported by the U.S. Department of Energy, Office of Science, Office of Nuclear Physics, from DE-SC0011090 and within the framework of the TMD Topical Collaboration.}


\begin{thebibliography}{0}%
\makeatletter
\providecommand \@ifxundefined [1]{%
 \@ifx{#1\undefined}
}%
\providecommand \@ifnum [1]{%
 \ifnum #1\expandafter \@firstoftwo
 \else \expandafter \@secondoftwo
 \fi
}%
\providecommand \@ifx [1]{%
 \ifx #1\expandafter \@firstoftwo
 \else \expandafter \@secondoftwo
 \fi
}%
\providecommand \natexlab [1]{#1}%
\providecommand \enquote  [1]{``#1''}%
\providecommand \bibnamefont  [1]{#1}%
\providecommand \bibfnamefont [1]{#1}%
\providecommand \citenamefont [1]{#1}%
\providecommand \href@noop [0]{\@secondoftwo}%
\providecommand \href [0]{\begingroup \@sanitize@url \@href}%
\providecommand \@href[1]{\@@startlink{#1}\@@href}%
\providecommand \@@href[1]{\endgroup#1\@@endlink}%
\providecommand \@sanitize@url [0]{\catcode `\\12\catcode `\$12\catcode
  `\&12\catcode `\#12\catcode `\^12\catcode `\_12\catcode `\%12\relax}%
\providecommand \@@startlink[1]{}%
\providecommand \@@endlink[0]{}%
\providecommand \url  [0]{\begingroup\@sanitize@url \@url }%
\providecommand \@url [1]{\endgroup\@href {#1}{\urlprefix }}%
\providecommand \urlprefix  [0]{URL }%
\providecommand \Eprint [0]{\href }%
\providecommand \doibase [0]{http://dx.doi.org/}%
\providecommand \selectlanguage [0]{\@gobble}%
\providecommand \bibinfo  [0]{\@secondoftwo}%
\providecommand \bibfield  [0]{\@secondoftwo}%
\providecommand \translation [1]{[#1]}%
\providecommand \BibitemOpen [0]{}%
\providecommand \bibitemStop [0]{}%
\providecommand \bibitemNoStop [0]{.\EOS\space}%
\providecommand \EOS [0]{\spacefactor3000\relax}%
\providecommand \BibitemShut  [1]{\csname bibitem#1\endcsname}%
\let\auto@bib@innerbib\@empty
\end{thebibliography}%


\begin{thebibliography}{99}


\bibitem{Ji:2013dva}
  X.~Ji,
  Parton Physics on a Euclidean Lattice,
  Phys.\ Rev.\ Lett.\  {\bf 110}, no. 26, 262002 (2013)
  [arXiv:1305.1539 [hep-ph]].


\bibitem{Xiong:2013bka} 
  X.~Xiong, X.~Ji, J.~H.~Zhang and Y.~Zhao,
  One-Loop Matching for Parton Distributions: Non-Singlet Case,
  Phys.\ Rev.\ D {\bf 90}, 014051 (2014)
  [arXiv:1310.7471 [hep-ph]].


\bibitem{Lin:2014zya}
  H.~-W.~Lin, J.~-W.~Chen, S.~D.~Cohen and X.~Ji,
  Flavor Structure of the Nucleon Sea from Lattice QCD,
  arXiv:1402.1462 [hep-ph].

\bibitem{Ma:2017pxb} 
  Y.~Q.~Ma and J.~W.~Qiu,
  Exploring hadrons' partonic structure using ab initio lattice QCD calculations,
  arXiv:1709.03018 [hep-ph].

\bibitem{Alexandrou:2015rja} 
  C.~Alexandrou, K.~Cichy, V.~Drach, E.~Garcia-Ramos, K.~Hadjiyiannakou, K.~Jansen, F.~Steffens and C.~Wiese,
  Lattice calculation of parton distributions,
  Phys.\ Rev.\ D {\bf 92}, 014502 (2015)
  doi:10.1103/PhysRevD.92.014502
  [arXiv:1504.07455 [hep-lat]].


\bibitem{Chen:2016utp} 
  J.~W.~Chen, S.~D.~Cohen, X.~Ji, H.~W.~Lin and J.~H.~Zhang,
  Nucleon Helicity and Transversity Parton Distributions from Lattice QCD,
  Nucl.\ Phys.\ B {\bf 911}, 246 (2016)
  doi:10.1016/j.nuclphysb.2016.07.033
  [arXiv:1603.06664 [hep-ph]].

\bibitem{Alexandrou:2016jqi} 
  C.~Alexandrou, K.~Cichy, M.~Constantinou, K.~Hadjiyiannakou, K.~Jansen, F.~Steffens and C.~Wiese,
  Updated Lattice Results for Parton Distributions,
  Phys.\ Rev.\ D {\bf 96}, no. 1, 014513 (2017)
  doi:10.1103/PhysRevD.96.014513
  [arXiv:1610.03689 [hep-lat]].


\bibitem{Chen:2017mzz} 
  J.~W.~Chen, T.~Ishikawa, L.~Jin, H.~W.~Lin, Y.~B.~Yang, J.~H.~Zhang and Y.~Zhao,
  Parton Distribution Function with Non-perturbative Renormalization from Lattice QCD,
  arXiv:1706.01295 [hep-lat].


\bibitem{Alexandrou:2017dzj} 
  C.~Alexandrou {\it et al.},
  Computation of parton distributions from the quasi-PDF approach at the physical point,
  arXiv:1710.06408 [hep-lat].



\bibitem{Ji:2014gla} 
  X.~Ji,
  Parton Physics from Large-Momentum Effective Field Theory,
  Sci.\ China Phys.\ Mech.\ Astron.\  {\bf 57}, no. 7, 1407 (2014)
  [arXiv:1404.6680 [hep-ph]].

\bibitem{Ji:2017rah} 
  X.~Ji, J.~H.~Zhang and Y.~Zhao,
  More On Large-Momentum Effective Theory Approach to Parton Physics,
  doi:10.1016/j.nuclphysb.2017.09.001
  arXiv:1706.07416 [hep-ph].

\bibitem{Zhang:2017bzy} 
  J.~H.~Zhang, J.~W.~Chen, X.~Ji, L.~Jin and H.~W.~Lin,
  Pion Distribution Amplitude from Lattice QCD,
  Phys.\ Rev.\ D {\bf 95}, no. 9, 094514 (2017)
  doi:10.1103/PhysRevD.95.094514
  [arXiv:1702.00008 [hep-lat]].

\bibitem{Lin:2017ani} 
  H.~W.~Lin, J.~W.~Chen, T.~Ishikawa and J.~H.~Zhang,
  Improved Parton Distribution Functions at Physical Pion Mass,
  arXiv:1708.05301 [hep-lat].


\bibitem{Radyushkin:2017cyf} 
  A.~V.~Radyushkin,
  Quasi-parton distribution functions, momentum distributions, and pseudo-parton distribution functions,
  Phys.\ Rev.\ D {\bf 96}, no. 3, 034025 (2017)
  doi:10.1103/PhysRevD.96.034025
  [arXiv:1705.01488 [hep-ph]].


\bibitem{Davoudi:2012ya} 
  Z.~Davoudi and M.~J.~Savage,
  Restoration of Rotational Symmetry in the Continuum Limit of Lattice Field Theories,
  Phys.\ Rev.\ D {\bf 86}, 054505 (2012)
  doi:10.1103/PhysRevD.86.054505
  [arXiv:1204.4146 [hep-lat]].

\bibitem{Detmold:2005gg} 
  W.~Detmold and C.~J.~D.~Lin,
  Deep-inelastic scattering and the operator product expansion in lattice QCD,
  Phys.\ Rev.\ D {\bf 73}, 014501 (2006)
  doi:10.1103/PhysRevD.73.014501
  [hep-lat/0507007].

\bibitem{Liu:1993cv} 
  K.~F.~Liu and S.~J.~Dong,
  Origin of difference between anti-d and anti-u partons in the nucleon,
  Phys.\ Rev.\ Lett.\  {\bf 72}, 1790 (1994)
  doi:10.1103/PhysRevLett.72.1790
  [hep-ph/9306299].

\bibitem{Liang:2017mye} 
  J.~Liang, K.~F.~Liu and Y.~B.~Yang,
  Lattice calculation of hadronic tensor of the nucleon,
  arXiv:1710.11145 [hep-lat].


\bibitem{Braun:2007wv} 
  V.~Braun and D.~Mueller,
  Exclusive processes in position space and the pion distribution amplitude,
  Eur.\ Phys.\ J.\ C {\bf 55}, 349 (2008)
  doi:10.1140/epjc/s10052-008-0608-4
  [arXiv:0709.1348 [hep-ph]].


\bibitem{Bali:2017gfr} 
  G.~S.~Bali {\it et al.},
  Pion distribution amplitude from Euclidean correlation functions,
  arXiv:1709.04325 [hep-lat].


\bibitem{Dorn:1986dt}
  H.~Dorn,
  Renormalization of Path Ordered Phase Factors and Related Hadron Operators in Gauge Field Theories,
  Fortsch.\ Phys.\  {\bf 34}, 11 (1986).
  doi:10.1002/prop.19860340104
 

\bibitem{Ishikawa:2016znu} 
  T.~Ishikawa, Y.~Q.~Ma, J.~W.~Qiu and S.~Yoshida,
  Practical quasi parton distribution functions,
  arXiv:1609.02018 [hep-lat].


\bibitem{Chen:2016fxx} 
  J.~W.~Chen, X.~Ji and J.~H.~Zhang,
  Improved quasi parton distribution through Wilson line renormalization,
  Nucl.\ Phys.\ B {\bf 915}, 1 (2017)
  doi:10.1016/j.nuclphysb.2016.12.004
  [arXiv:1609.08102 [hep-ph]].

\bibitem{Ji:2017oey} 
  X.~Ji, J.~H.~Zhang and Y.~Zhao,
  Renormalization in Large Momentum Effective Theory of Parton Physics,
  arXiv:1706.08962 [hep-ph].

\bibitem{Ishikawa:2017faj} 
  T.~Ishikawa, Y.~Q.~Ma, J.~W.~Qiu and S.~Yoshida,
  On the Renormalizability of Quasi Parton Distribution Functions,
  arXiv:1707.03107 [hep-ph].

\bibitem{Green:2017xeu} 
  J.~Green, K.~Jansen and F.~Steffens,
  Nonperturbative renormalization of nonlocal quark bilinears for quasi-PDFs on the lattice using an auxiliary field,
  arXiv:1707.07152 [hep-lat].

\bibitem{Nocera:2017war} 
  E.~R.~Nocera, H.~W.~Lin, F.~Olness, K.~Orginos and J.~Rojo,
  The PDFLattice2017 workshop: a summary report,
  arXiv:1709.01511 [hep-ph].

\bibitem{Stewart:2017tvs} 
  I.~W.~Stewart and Y.~Zhao,
  Matching the Quasi Parton Distribution in a Momentum Subtraction Scheme,
  arXiv:1709.04933 [hep-ph].


\bibitem{Accardi:2016qay} 
  A.~Accardi, L.~T.~Brady, W.~Melnitchouk, J.~F.~Owens and N.~Sato,
  Constraints on large-$x$ parton distributions from new weak boson production and deep-inelastic scattering data,
  Phys.\ Rev.\ D {\bf 93}, no. 11, 114017 (2016)
  doi:10.1103/PhysRevD.93.114017
  [arXiv:1602.03154 [hep-ph]].

\bibitem{Monahan:2016bvm} 
  C.~Monahan and K.~Orginos,
  Quasi parton distributions and the gradient flow,
  JHEP {\bf 1703}, 116 (2017)
  doi:10.1007/JHEP03(2017)116
  [arXiv:1612.01584 [hep-lat]].


\end{thebibliography}
\end{document}